Review Article

# Elucidation of role of graphene in catalytic designs for electroreduction of oxygen

(*Short title:* role of graphene component in electrocatalysts for reduction of oxygen)

*Pawel J. Kulesza*[a]*, Jerzy K. Zak*[a,b]*, Iwona A. Rutkowska*[a]*, Beata Dembinska*[a]*, Sylwia Zoladek*[a]*, Krzysztof Miecznikowski*[a]*, Enrico Negro*[c]*, Vito Di Noto*[c]*, Piotr Zelenay*[a,d]

[a]Faculty of Chemistry, University of Warsaw, Pasteura 1, 02-093 Warsaw, Poland

[b]Institute of Physical Chemistry and Technology of Polymers, Faculty of Chemistry, Silesian University of Technology, Strzody 9, 44-100 Gliwice, Poland

[c]Department of Industrial Engineering, Università degli Studi di Padova in Department of Chemical Sciences, Via Marzolo 1, 35131 Padova, Italy

[d]Los Alamos National Laboratory, Materials Physics and Applications, Los Alamos, New Mexico 87545, U.S.A

*Corresponding Author: P.J. Kulesza (pkulesza@chem.uw.edu.pl)

**Keywords:** graphene-based catalysts; hybrid materials; doping; modification of surfaces; metal nanoparticles; activating interactions; oxygen reduction

Graphene is, in principle, a promising material for consideration as component (support, active site) of electrocatalytic materials, particularly with respect to reduction of oxygen, an electrode reaction of importance to low-temperature fuel cell technology. Different concepts of utilization, including nanostructuring, doping, admixing, preconditioning, modification or functionalization of various graphene-based systems for catalytic electroreduction of oxygen are elucidated, as well as important strategies to enhance the systems' overall activity and stability are discussed.

**Highlights**

- Doping of graphene by replacement of lattice carbons with heteroatoms such as nitrogen influences local charge densities and affects adsorptive interactions with oxygen.

- Interfacial coordination of certain transition metals and their oxides to heteroatom sites further enhances the electrocatalytic activity, particularly in alkaline solutions.

- Reduced graphene oxide can be used to support noble metal nanoparticles for oxygen reduction in acid media.

- Typically the defect-rich graphene supports exhibit higher activity.

- Hybrid or admixed graphene-based catalysts with different carbons often exhibit increased stability and  catalytic activity

**Introduction**

There has been continuous interest in development of electrocatalytic systems for oxygen reduction reaction (ORR), particularly with respect to potential applications in low-temperature fuel cells [1-10]. In addition to such important issues as improvement of stability and activity, a recent trend to replace highly expensive platinum-rich materials with alternative electrocatalysts should be noted. In this respect, special attention has been devoted to noble-metal-free carbon-based (e.g. heteroatom-doped, iron or cobalt modified, totally-metal-free functionalized or derivatized) nanomaterials including graphene-type systems [11-17]. Despite intensive research in the area, there are many fundamental problems to be resolved, and, therefore, the practical ORR catalysts are still based on platinum. In this respect, there is a need of better utilization of catalytic sites and significant lowering of the noble metal loadings.

The optimum electrocatalytic materials should be characterized by large electrochemically active surface areas and contain active sites dispersed on suitable supports. While exhibiting long-term stability, a useful support should prevent agglomeration of catalytic centers, exhibit activating interactions with them, facilitate oxygen mass transfer and water removal, and assure good electrical conductivity to active sites at the electrocatalytic interface. Because of unique thermal, mechanical and electrical properties coupled with the high specific surface area, various graphene-based electrocatalysts have been recently proposed and characterized [11-19]. In this respect, many observations require summarizing and clarification.

**Heteroatom-doped graphene based electrocatalysts**

It has been established that doping of the originally inert graphene may lead to formation of catalytically active sites by replacement of the lattice carbons with heteroatoms [20,21]. Because N-doped graphene catalysts feature graphitic, pyridinic or pyrrolic nitrogens differing in chemical identities of the heteroatom and the local charge densities and the spin density redistribution on the adjacent carbon atoms, there should differences in the oxygen adsorption, activation and its reductive activity.

In general, fairly high level of activity toward ORR has been demonstrated for the N-doped graphene catalysts in alkaline media, contrary to the relatively poorer performance in acidic environment. Further enhancement of activity of the N-doped systems can be achieved through coordination of such transition metals as Fe or Co to the N-sites [22]. The resulting electrocatalysts have occurred to be active for ORR in primarily alkaline but also acid media.

Not only transition metals but also their oxides can be considered for ORR together with graphene. Among important issues are their mixed valence character, presence of donor-acceptor chemisorptions sites, and the ability to adsorb reversibly oxygen. Care must be exercised to minimize ohmic drops at the electrocatalytic interface. A hybrid system composed of $Co_3O_4$ nanoparticles deposited onto graphene substrate has been demonstrated to exhibit a synergistic effect (relative to bare graphene and $Co_3O_4$ itself) and act as effective electrocatalyst for ORR [23]. Presumably due to specific metal-nitrogen interactions, application of N-doped graphene provides a better distribution of the metal oxide nanoparticles [24].

**Graphene-based carriers for noble metal nanoparticles**

The undoped-graphene-based catalytic systems, namely without any heteroatom in its structure, were also reported [25,26]. A representative example concerned the well-behaved (fairly high current densities, low over-potentials) system composed of graphene quantum dots supported onto graphene nanoribbons at which the four-electron ORR was operative in alkaline medium [25]. The edge-rich dopant-free graphene-based system showed high ORR activity as well [26]. The actual activity of bare graphene type systems is a question of dispute. Certainly, the surface physicochemical identity (e.g. degree of oxidation and distortion) and possible contaminations (e.g. with Mn), in addition to practical durability, could be issues as well. Intentional decoration with noble metal (mostly Pt) nanoparticles at as low as possible loadings seems to be a reasonable solution.

Despite serious limitations (high cost, possibility of corrosion and deactivation), platinum based systems are so far the most established and the best known electrocatalysts for ORR [1-4]. But the effectiveness and long-term performance of platinum sites would be largely dependent on a choice of support. Various carbons are considered as substrates but, in addition to their stability and durability of the Pt adherence, certain strategies to reduce loadings of noble metal at the interface have been proposed. They include improvement of the dispersion by reducing the metal particle size and proper supporting with strong attraction to the substrate. An ideal support would provide high conductivity and large surface area, in addition to stabilization and potential activation or synergistic effects. Furthermore, the sufficiently porous and rigid supports would permit utilization of alloyed nanoparticles. Finally, the specific noble metal – carbon lattice interactions have been postulated. But their nature and consequences require experimental verification.

Graphene-type supports are certainly capable of anchoring noble metal nanoparticles [11,13,15]. But graphene can exist in different forms. A single sheet of $sp^2$ hybridized carbon atoms is usually obtained from bulk graphite material by mechanical exfoliation or via chemical procedures. The latter approach may lead to additional doping of the carbon lattice or introduction of functional groups. The chemical oxidation (to graphene oxide) results in the increased hydrophilicity due to formation of oxygen-containing (e.g. carboxyl or epoxide) groups. While high porosity of the oxidized form (graphene oxide) should be appreciated when it comes to its potential application as support for catalytic nanoparticles, the oxidized system's overall stability (resistance to oxidative corrosion) and conductivity are decreased. A reduction step, that is typically performed chemically and is irreversible, produces so called reduced graphene oxide (RGO). In comparison to the oxidized form, the latter system is more hydrophobic but it still contains hydrophilic domains [27]. Among important diagnostic experiments, spectroscopic monitoring of surface oxygen-containing functional groups, such as –C-OH, -COOH, -C=O and -C-O-C- (using FTIR), and of relative intensities of G and D bands at about 1600 and 1350 cm$^{-1}$ (using Raman) should be mentioned. Typically RGO surfaces are characterized by the low degree of organization of graphitic structure and by the presence of defects. When these interfacial features are combined with some hydrophilicity, the increased adsorptive capabilities and enhanced catalytic activity of RGO can be envisioned. It is also noteworthy that RGO is likely to act as robust and conductive catalytic material or carrier [27,28]. Thus RGO seems to be the optimum form of graphene for electrocatalytic applications, including oxygen reduction.

A classical procedure for deposition of Pt nanoparticles onto graphene sheets involved reduction (using NaBH$_4$) of H$_2$PtCl$_6$ in the suspension of graphene oxide (GO) [28]. The resulting partially reduced GO-Pt catalyst was subjected to further reduction with hydrazine at elevated temperature (300 $^0$C). The observed enhancement of the electrochemically active surface area was correlated with a spatial morphology of the catalytic film in which Pt nanoparticles (anchored to single graphene sheets) induced separation between graphene flakes. Under such conditions, transport of reactants (oxygen, water) was feasible. When the catalyst was tested as cathode material in low-temperature polymer membrane fuel cell, a maximum power of 161 mW cm$^{-2}$ was achieved (relative to 96 mW cm$^{-2}$ obtained with use of an unsupported Pt nanoparticles).

Graphene support could facilitate electron transport through the system's basal planes to edges. Faster electron transfers at edges resulted in lower contact resistances when particles are deposited therein. Indeed, the ball-milled graphene supports exhibited higher activity than those based on graphene flakes during ORR [29]. Obviously the ball-milling procedure tended to produce the defect-rich oxidized surface-functionalities. Their presence was of particular importance during experiments in alkaline medium. A synthetic CVD methodology was explored for fabrication of electrodes of controlled morphology [30]. Such systems permitted controlled arrangement of deposited metal nanoparticles.

**Importance of metal-support interactions**

A unique structure composed of web-like horizontally-aligned carbon nanotubes and graphene was synthesized in fluidized-bed microreactor [31]. Here carbon nanotubes were attached mainly to the edges of graphene thus making the hybrid structure not only highly porous (with specific surface area on the level of 863 $m^2$ $g^{-1}$) but also capable of preventing graphene stacking and creating numerous sites for the deposition of Pt nanocrystals. It is noteworthy that the system exhibited fairly high activity at low Pt loadings thus promising with respect to reducing the cost of fuel cell electrocatalyst.

An ideal support (including graphene-type carriers) would interact beneficially with the metal centers or even with reactants including the reaction intermediates. This issue should be taken into account together with proper morphology, functionalization or related reactivity, stability toward chemical and electrochemical attack, and structure of the interface. To get insight into nature of the enhancement effect observed at the above-mentioned web-like horizontally-aligned hybrid electrocatalyts composed of carbon nanotubes (CNTs) and graphene [31], a computational model was built and the density functional theory (DFT) calculations were considered for defective graphene samples with the abundant active edge sites and 13-atom-Pt clusters around. For platinum existing in such environment, a significant downshift of d-band center was postulated. Typically, the 3d electrons existing close to the Fermi level are characterized by strong binding energy, whereas those existing further away from the Fermi level have weaker binding energy. By supporting Pt clusters onto defective graphene, the binding strength of oxygen (to platinum surface) was possibly reduced to the level facilitating activation of the $O_2$ molecule and promoting the ORR kinetics. The results of X-ray photoelectron spectroscopic (XPS) experiments confirmed presence of Pt, C and O; furthermore, it was apparent from the Pt4f XPS spectra that the Pt surface was mainly in the metallic state. Upon comparison of the Pt4f binding energies for platinum deposited on the conventional carbon black (20wt% Pt on Vulcan XC-72) and on the hybrid CNT/graphene supports, a positive shift of ca. 0.5 eV was observed in the latter case. This result implied relatively stronger interactions between Pt nanocrystals and the hybrid support with large population of edge sites. Here the presence of graphitic structure and the existence of defects and the largely exposed edge sites was confirmed upon deconvolution of C1s peaks. Thus the performance was strongly dependent on the graphene morphology and formation of hybrid carriers by combination or alignment with carbon nanotubes.

Utilization of catalytic nanoparticles based on alloys of platinum often results in the increased electrocatalytic activity during ORR [32]. The electronic interactions between Pt and the other alloy-forming metal would lead to changes in the interfacial electronic densities of active platinum. Furthermore, improved stability of the alloyed catalyst has been postulated.

For example, to prepare uniformly-distributed $Pt_2Pd$ alloyed-nanoparticles embedded in nitrogen-rich graphene nanopores, platinum phthalocyanine (PtPc) and palladium phthalocyanine (PdPc) were used (following calcination at high temperatures) as precursors of both alloy-forming metals as well as the source of abundant nitrogen sites or dopants within the graphene skeleton [33]. Strong interactions between $Pt_2Pd$ alloy nanoparticles and N-derivatized graphene units (within nanopores) were postulated to explain the observed enhancement effects during hydrogen evolution and oxygen reduction reactions. Special attention was paid to the N1*s* XPS spectra which were deconvoluted to five peaks: pyridinic-N, at 398.8 eV; N-M (M-metal), at 399.4 eV; pyrrolic-N, at 400 eV; graphitic-N, at 401.1 eV; and oxidized-N, at 402.5 eV. The appearance of the peak characteristic of N-M implied possibility of strong interactions of the $Pt_2Pd$ alloy within N-doped nanopores. These features were correlated with the increased electrocatalytic activity (ORR in 0.1 mol $dm^{-3}$ KOH) and the system's good stability preventing agglomeration of catalytic $Pt_2Pd$ nanoparticles. Indeed, the pyridinic-N and pyrrolic-N could work as metal coordination sites because of the capability of donating lone-pair electrons. The XPS spectra taken in the regions characteristic of platinum and palladium implied existence (in both cases) of their metallic and oxidized forms. The feasibility of stronger adherence of $Pt_2Pd$ nanoparticles to N-derivatized, rather than undoped, graphene sites was supported with DFT calculations. On the whole, the performance of N-doped-graphene-supported $Pt_2Pd$ nanoparticles was competitive to (if not superior than) that of the commercial Vulcan-supported Pt during ORR. Thus it can be rationalized that N-doping of porous graphene supports (for noble metal catalytic nanoparticles) should generally improve activity of the graphene-based electrocatalytic systems in $O_2$-saturated alkaline solutions.

The other electrocatalytic system utilized RGO onto which gold nanoparticles were introduced [11]. Ultra-thin adsorbate-type layers of polyoxometallates (namely Keggin-type phosphododecamolybdates) were demonstrated to adsorb strongly on both metal (Au) and RGO surfaces thus stabilizing Au nanoparticles (as capping ligands), linking them to the RGO supports and facilitating their dispersion. During experiments in alkaline medium polyoxometallate adsorbates were removed to yield bare highly active Au nanostructures immobilized onto RGO supports. The study clearly demonstrated that the chemically-reduced graphene-oxide (with structural defects existing on poorly organized graphitic structure) acted as a robust and activating support for dispersed gold nanoparticles during electrocatalytic reduction of oxygen in alkaline medium (0.1 mol dm$^{-3}$ KOH). For the same loadings (30 μg cm$^{-2}$) and comparable morphologies of catalytic gold nanoparticles, the system utilizing RGO support produced, relative to the unsupported Au and the Vulcan-carbon-black supported Au-catalysts, lower amounts of the undesirable $HO_2^-$ intermediate. Furthermore, application of RGO as support for Au evidently affected the onset potential for ORR and shifted it toward more positive values (0.9 V). The observed synergism could be interpreted in terms of activating interactions between catalytic Au nanoparticles and graphene carrier [34-36] leading to lowering of the dissociation energy of $O_2$ molecule adsorbed on Au-metal in presence of RGO or to reduction of the stability of $HO_2^-$-intermediate species. Furthermore, such supports composed of Au nanoparticles immobilized on RGO were also demonstrated to work as active carriers for dispersed Pt nanoparticles during ORR both in alkaline [11] and acid (0.5 mol dm$^{-3}$ $H_2SO_4$) media [37]. In the latter case, the ability of RGO-supported Au to induce decomposition of hydrogen peroxide intermediate should be mentioned. Introduction and dispersion of Au nanoparticles improved charge distribution within the catalytic system and facilitated exfoliation and separation of graphene flakes to assure unimpeded flux of reactants. Finally, the system composed of RGO-supported Ir nanostructures admixed with carbon-nanotube-supported Pt nanoparticles (at low metal loadings, 1.5 and 15 μg cm$^{-2}$, respectively) yields highly active electrocatalytic system for ORR in acid medium with low production of $H_2O_2$ [37].

High activity of RGO-supported Au alloyed with Pd has also been demonstrated for ORR in alkaline medium [38]. It has been apparent from the XPS data (C1s spectrum) that, following solvothermal formation of monodispersed bimetallic AuPd nanoparticles within RGO, the C-C bonds, rather than the C-O bonds (oxygenous groups), are becoming predominant on the graphene surfaces. Furthermore, the Au4f XPS signals of the RGO-supported Au-Pd nanoparticles can only be deconvoluted into two peaks at ca. 87.02 and 83.3 eV, thus indicating the existence of metallic $Au^0$ as the main species (known to be reactive toward $O_2$-electroreduction in $O_2$-sataruted 0.1 mol $dm^{-3}$ KOH). The system's electrocatalytic activity has been also promoted by the presence of $Pd^0$ and $Pd^{2+}$ sites. Under such conditions the electron transfer numbers can reach 3.95 during ORR (i.e. very close to theoretical 4-electron pathway consistent with the direct reduction to $H_2O$).

The theoretical approach based on a first principle calculations using the Vienna *ab initio* simulation package aiming at addressing interactions between the graphene-substrate and Pd nanoparticles [39] should be mentioned here. The choice of Pd can be rationalized as follows: Pd is a Pt-like metal, and its activity is enhanced toward ORR upon Pd alloying or supporting on certain substrates. For a model $Pd_{13}$ nanoparticle attached to a single vacancy graphene, the binding energy has been found to be at ca -6.10 eV as a result of hybridization of Pd particles with the $sp^2$ dangling bonds at the defect sites. The interaction should result in a shift of the d-band center for the Fermi level from -1.02 to -1.45 eV. If the graphene single vacancy is doped with B or N, an additional shift of the d-band center should occur; it is capable of changing the system's activity or affinity toward O, OH, and OOH adsorption. Based on calculations, the adsorption energies of O, OH, and OOH are lowered from -4.78, -4.38 and -1.56 eV for the freestanding $Pd_{13}$ nanoparticle to the following valuses, -4.57, -2.66 and -1.39 eV, that are characteristic of $Pd_{13}$ on a single vacancy of doped graphene. The result is consistent with experimental observations implying that defects in graphene substrates are capable of effectively stabilizing Pd nanoparticles. Under such conditions, the adsorption interactions between O-containing species and Pd nanoparticles are likely to be weaker too.

**Hybrid systems**

Surprisingly high catalytic activity toward ORR (in alkaline medium) is exhibited by the hybrid system formed by $Co_3O_4$ nanocrystals grown on the mildly-reduced RGO [40]. It is noteworthy that $Co_3O_4$, graphene oxide and RGO exhibit low catalytic activity when considered as single components under analogous experimental conditions. The catalyst's performance was further improved by N-doping through addition of $NH_4OH$ during the synthesis. It has been apparent from rotating disk voltammetric measurements that the latter system has behaved in a manner analogous to the commercial Pt/C catalyst. The result is consistent with the involvement of almost 4-electrons during the reduction process in the potential range from 0.6 to 0.75 V (vs RHE). The half-wave potential has been determined to be equal to 0.83 V at 1600 rpm (0.86 V has been found for Pt/C). The rotating ring-disk electrode (RRDE) measurements have been employed to monitor the formation of peroxide type species ($HO_2-$) during the ORR process and have yielded values not exceeding 6% over the potential range of 0.45-0.8 V with the electron transfer number of 3.9. On the contrary, the bare ($Co_3O_4$-free) N-doped RGO support has exhibited low ORR activity involving predominantly only 2 electrons.

It is generally accepted that N-doping of graphene is fairly simple and facilitates its usefulness as active material or substrate for ORR catalysts in alkaline media. The experimental results supported with calculations indicate that N-doped sites in graphene structure form stable centers for attachment of noble and non-precious metal nanoparticles. In this respect, both active sites at graphene edges and in the basal plane are of importance. Doping of the graphene basal plane with Co–N has been addressed using the 1st-principle calculations and *ab initio* molecular dynamics simulations with respect to possible ORR mechanisms [41]. Theoretical considerations of geometric structures, kinetics and reaction pathways feasible on N−G, Co−G, and Co−$N_4$−G surfaces have also been addressed. The results demonstrate that simple N-doping of graphene is not enough for effective breaking the O−O bond in $O_2$ or HOO*. The situation changes following Co-doping of graphene, particularly when the Co-$N_4$ complex is introduced to the basal plane: it activates of $O_2$ dissociation and facilitates the H₂O desorption. Consequently, the ORR proceeds on the Co−N$_4$−functionalized graphene surfaces with relatively lower overpotentials. But further improvement of the electrocatalytic activity requires optimization of the coupling between transition-metal and nonmetallic dopants in the graphene basal plane. For example, a hybrid-assembly of graphene oxide, o-phthalonitrile and cobalt acetate has been converted (following thermal treatment under nitrogen flow) to the macroporous framework of N-doped graphene sheets with uniformly deposited Co nanoparticles [42]. Upon subjecting to pyrolysis at 600 °C, the resulting catalyst has exhibited electrocatalytic behavior comparable to that known for the commercial Pt/C catalyst in alkaline electrolyte. Another powerful ORR catalyst composed of N-doped graphene and Co nanoparticles encased in N-doped graphitic shells has been fabricated upon subjecting a mixture of sucrose, urea and cobalt nitrate to one-step pyrolysis [43]. Here the 4-electron ORR pathway has been dominating in alkaline media. Using the XPS data, it can be stated that the graphitic lattice is not perfect and reveals the presence of doped or modified heteroatoms (N and O). The N1s XPS spectrum implies existence of pyridinic (398.7 eV), graphitic (401.1 eV), pyrrolic (399.8 eV) and oxidized (402.1 eV) N-species. The fact, that graphitic and pyridinic N-species have been in majority, facilitates the electrocatalytic ORR. After deconvolution of the O1s spectrum, the existence of CoO can also be postulated. The presence of metallic cobalt has been apparent form XRD analysis (but not XPS).

A unique electrocatalytic system has been based on nanoporous N-doped carbon/graphene nano-sandwiches [44]. Here the zeolitic-imidazolate-framework nanocrystalline layers of ca. 20 nm diameter have been initially grown on both sides of graphene oxide to obtain the sandwich-type structure. After carbonization, acid etching, and N-doping, catalytic surfaces of high specific surface area (1170 m$^2$ g$^{-1}$) have been obtained. Using such catalysts, the onset potential for ORR (in 0.1 mol dm$^{-3}$ KOH) has been fairly positive (0.92 V vs. RHE); furthermore, fairly large limiting current densities (on the level of 5.2 mA cm$^{-2}$ at 0.60 V) have been observed under rotating disk conditions (1600 rpm).

Further functionalized hybrid systems can exhibit even better performance. For example, nanonporous Co-N$_x$/carbon nanosheets can be prepared using not only cobalt nitrate but also zinc nitrate as co-metal sources. The resulting catalytic system is characterized by highly positive onset potential for oxygen reduction (0.96 V) competitive to that characteristic of the commercial Pt/C studied under analogous conditions (0.94 V). It is noteworthy that the latter system exhibits reasonable ORR electrocatalytic activity in acidic medium (with the onset potential of 0.85 V; nevertheless, somewhat lower than that for commercial Pt/C, 0.93 V). Such features as stability and good methanol tolerance should also be mentioned here.

Multifunctional catalysts, capable of inducing such processes as oxygen evolution, oxygen reduction or hydrogen (in alkaline media), have been obtained by integrating certain metals (Ni or Co) with their oxides (NiO or CoO) and dispersing them onto nitrogen-doped RGO [45]. With respect to ORR, the cobalt-based system seems to exhibit the highest electrocatalytic activity, and its potential applicability in primary zinc-air batteries (comparable performance to the Pt/C or Pt/C+IrO$_2$ cathodes) has been demonstrated.

Graphene-supported iron nanostructures could also exhibit remarkable electrocatalytic activity toward ORR in alkaline medium. A representative example utilizes iron nanoparticles (5.0 wt% Fe) generated within nitrogen moieties following thermal decomposition (N content from 4.0% to 12.0%, depending on temperature) of graphitic carbon nitride immobilized on graphene sheets containing iron salts (*e.g.*, FeCl$_3$) [46]. In addition to the performance of this nitride-graphene composite resembling electrocatalytic behavior of the commercially available Pt-based electrodes (30 wt % Pt on Vulcan XC72), the system's long-term stability and tolerance to methanol crossover effect should be mentioned. In another example of the efficient catalyst for oxygen electroreduction, both Fe and Co have been coordinated to N-doped active edge-sites created by controlled oxidative etching of porous graphene. Here the coexistence of N-coordinated bimetallic (Co/Fe) centers with nitrogen sites locked at the pore openings tends to reduce substantially the overpotential for ORR in alkaline medium [46,47]. Because of very good stability, the catalytic system has been considered as cathode material in the anion exchange membrane fuel cell: here a power density of ca. 35 mW cm$^{-2}$ has been achieved relative to 60 mW cm$^{-2}$ displayed by the Pt-based system. Finally, the concept of dual-doping of there-dimensional (3-D) graphene with both nitrogen and sulfur to produce a substrate for catalytic $CoFe_2O_4$ nanoparticles should be mentioned here [48]. An important feature of the graphene with 3-D hierarchical porous structure is that it provides not only a good conducting network for electrons but also promotes the efficient mass transport of the reactant ($O_2$, $H_2O$) molecules and electrolyte species during the redox processes. Under such conditions, more catalytic sites seem to be utilized because stacking of graphene layers seems to be suppressed.

Preparation of hierarchical, particularly noble-metal free, catalytic systems is of particular importance for the ORR process. Core-shell-type Pt-free highly-active electrocatalysts have been recently proposed [49]. They utilize metal ($FeSn_{0.5}$) based nanoparticles embedded in a carbon nitride shell templated on the graphene nanoplatelet core support. Here, instead of conventional metal, metal oxide or alloyed metal nanoparticles, sub-nanometric metal (Fe and Sn) clusters, which are bound within coordination nests formed by carbon and nitrogen ligands existing on surfaces of the carbon nitride shells, have been demonstrated to act the catalytically active sites during ORR in alkaline medium. Diagnostic experiments with use of RRDE voltammetry and gas-diffusion electrode measurements point on ORR overpotential of only *ca*. 70 mV higher than that observed at the XC-72R carbon black supported platinum. Among important issues are high stability, selectivity towards the

4-electron reduction of oxygen to water, high tolerance to methanol poisoning, and possibility of application in anion-exchange membrane fuel cells.

**Stabilization approaches**

Practical ORR electrocatalysts are often exposed to potentials higher than 0.9 V vs. RHE. During long-term operation under such conditions, oxidative corrosion of virtually all carbons, including graphene-type mterials, takes place. Degradation of carbon supports results typically in loss of catalytic centers or in their poisoning (e.g. in a case of Pt with -CO, -CHO or -COH adsorbates). In an attempt to produce a system of improved durability, graphene nanoplatelets, which were produced by exfoliating the intercalated graphite through modification with (poly(diallyldimethylammonium chloride), PDDA, were demonstrated to act as an alternative support material for Pt nanoparticles [50]. The stabilization effect reflected apparently the intrinsic high graphitization degree of graphene nanoplatelets and their enhanced interactions with Pt. In another example of the highly durable ORR catalyst, a hybrid composite material was prepared by careful mixing carbon black with RGO-supported Pt [51]. Here by inserting carbon black nanostructures between the platinized RGO, its durability was significantly improved. It was postulated that the two-dimensional RGO functioned as a "mesh" preventing leaching of dissolved Pt species into the electrolyte, whereas carbon black nanostructures could serve as active sites for recapture or renucleation of small Pt clusters. Similar stabilization effects were postulated for the platinized sandwich structure (support) in which VulcanXC-72 is placed in between graphene nanosheets [52]. Stabilities of different supports (for Pt nanoparticles), such as carbon black, reduced graphene oxide, and multiwalled carbon nanotubes were compared under harsh electrochemical conditions (0.1 M $HClO_4$) [53]. It was found that while the latter substrate was the most stable, the first one tended to degrade the most readily. Thus RGO could be considered as the moderately stable substrate. Further research is needed along this line because stability reflects many experimental parameters including the material's porosity and the degree of formation of the hydrogen peroxide intermediate.

**Conclusions and perspectives**

Graphene-based materials can be considered as catalytic components but their chemical identity largely depends on expected function (support or active sites) and reaction conditions (pH, type of electrolyte etc.). Of particular interest to studies in alkaline media are functionalized, particularly N-doped graphene catalysts in which the lattice carbons are replaced with heteroatoms. Coordination of such surface sites with transition metals (e.g. Fe or Co) would often lead to enhancement of the electrocatalytic activity (mostly in alkaline but also in acid media). A concept of preparation of hierarchical core-shell-type systems, in which non-noble metal nanoparticles are embedded in carbon nitride shells covering graphene-nanoplatelet core –supports, can be viewed as a promising approach to stabilization, activation and nanostructuring of graphene-based ORR catalysts. Also hybrid structures of graphene admixed with different carbons are useful materials because of improved mass transport of reactants, stability and catalytic efficiency.

Reduced graphene oxides with interfacial defects, moderate hydrophilicity, low degree of organization and porosity can be successfully used as supports capable of anchoring noble metal nanoparticles (e.g. Pt, Pd, Au, Ag, bimetallic alloys) for ORR in acid media. Among important issues are specific metal-support activating interactions. In this respect, further research aiming at derivatization of graphene surfaces, e.g. with hydroxyl- group-rich metal oxides such $TiO_2$, $ZrO_2$ or $SiO_2$, is expected.

More attention has to be paid to requirements coming from the ORR pathways, i.e. to mechanistic and kinetic details. Formation of hydrogen peroxide type intermediates would have to be further decreased in practical systems. In other words, the graphene-based systems are likely to multi-component and multi-functional with distinct catalytic sites including those capable of inducing decomposition of $H_2O_2$.

**Acknowledgements**

We acknowledge the European Commission through the Graphene Flagship – Core 1 project [Grant number GA-696656] and Maestro Project [2012/04/A/ST4/00287 (National Science Center, Poland)].